# Time and Memory Efficient Lempel-Ziv Compression Using Suffix Arrays

Artur J. Ferreira[1,3,4], Arlindo L. Oliveira[2,4], Mário A. T. Figueiredo[3,4]

[1]*Instituto Superior de Engenharia de Lisboa, Lisboa, PORTUGAL*
[2]*Instituto de Engenharia de Sistemas e Computadores, Lisboa, PORTUGAL*
[3]*Instituto de Telecomunicações, Lisboa, PORTUGAL*
[4]*Instituto Superior Técnico, Lisboa, PORTUGAL*
Contact email: `arturj@cc.isel.ipl.pt`

## Abstract

The well-known dictionary-based algorithms of the Lempel-Ziv (LZ) 77 family are the basis of several universal lossless compression techniques. These algorithms are asymmetric regarding encoding/decoding time and memory requirements, with the former being much more demanding, since it involves repeated pattern searching. In the past years, considerable attention has been devoted to the problem of finding efficient data structures to support these searches, aiming at optimizing the encoders in terms of speed and memory. Hash tables, binary search trees and suffix trees have been widely used for this purpose, as they allow fast search at the expense of memory.

Some recent research has focused on *suffix arrays* (SA), due to their low memory requirements and linear construction algorithms. Previous work has shown how the LZ77 decomposition can be computed using a single SA or an SA with an auxiliary array with the longest common prefix information. The SA-based algorithms use less memory than the tree-based encoders, allocating the strictly necessary amount of memory, regardless of the contents of the text to search/encode.

In this paper, we improve on previous work by proposing faster SA-based algorithms for LZ77 encoding and sub-string search, keeping their low memory requirements. For some compression settings, on a large set of benchmark files, our low-memory SA-based encoders are also faster than tree-based encoders. This provides time *and* memory efficient LZ77 encoding, being a possible replacement for trees on well known encoders like LZMA. Our algorithm is also suited for text classification, because it provides a compact way to describe text in a bag-of-words representation, as well as a fast indexing mechanism that allows to quickly find all the sets of words that start with a given symbol, over a static dictionary.

## 1 Introduction

The Lempel-Ziv 77 (LZ77) and its variant Lempel-Ziv-Storer-Szymanski (LZSS) [14, 16, 19] lossless compression algorithms are the basis of a wide variety of universal source coders, such as GZip, WinZip, PkZip, WinRar, and 7-Zip, among others. Those algorithms are asymmetric in terms of time and memory requirements, with encoding being much more demanding than decoding.

The LZ-based encoders use efficient data structures, like *binary trees* (BT) [6, 11], *suffix trees* (ST) [5, 7, 9, 13, 17] and hash tables, thus allowing fast search at the expense of higher memory requirement. The use of a Bayer-tree along with special binary searches, on a sorted sliding window, to speedup the encoding procedure, has been addressed [6]. *Suffix arrays* (SA) [7, 10, 15], due to their simplicity, space efficiency, and linear time construction algorithms [8, 12, 18] have been a focus of research; for instance, the linear time SA construction algorithm *suffix array induced sorting* (SA-IS) has been recently proposed [12].

SA have been used in encoding data with anti-dictionaries [4], to find repeating sub-sequences [1] for data deduplication, among other applications. Recently, space-efficient algorithms for computing the LZ77 factorization of a string, based on SA and auxiliary arrays, have been proposed to replace trees [2, 3]. These SA-based encoders require less memory than ST-based encoder, with some penalty on the encoding time, for roughly the same compression ratio. The amount of memory for the SA-based encoder is constant, independent of the contents of the sequence to encode, as opposed to tree-based encoders in which has to be allocated a maximum amount of memory.

In this paper, we improve on previous approaches [2, 3], proposing faster SA-based algorithms for LZ77/LZSS encoding, without requiring any modifications on the decoder side. These low-memory encoders are faster than the tree-based ones, like 7-Zip, being close to GZip in encoding time on several standard benchmark files.

The rest of this paper is organized as follows. Section 2 presents the basic concepts of LZ77/LZSS encoding using suffix arrays. Section 3 describes our proposed algorithm. The experimental results are discussed in Section 4 and some concluding remarks are made in Section 5.

## 2 Lempel-Ziv Compression using Suffix Arrays

The LZ77 and LZSS [14, 16, 19] lossless compression techniques use a sliding window over the sequence of symbols to be encoded, which has two sub-windows: the *dictionary* (holding symbols already encoded) and the *look-ahead-buffer* (LAB, containing the next symbols to be encoded). As the string in the LAB is encoded, the window slides to include it in the dictionary (this string is said to *slide in*); consequently, the symbols at the far end of the dictionary are dropped (*slide out*).

At each step of the LZ77/LZSS encoding algorithm, the longest prefix of the LAB which can be found anywhere in the dictionary is determined and its position stored. For these two algorithms, encoding of a string consists in describing it by a token. The LZ77 token is a triplet of fields, (*pos, len, sym*), with the following meanings:

- *pos* - location of the longest prefix of the LAB found in the current dictionary; this field uses $\log_2(|\text{dictionary}|)$ bits, where |dictionary| denotes the length (number of bytes) of the dictionary;

- *len* - length of the matched string; this requires $\log_2(|\text{LAB}|)$ bits;

- *sym* - the first symbol in the LAB that does not belong to the matched string (*i.e.*, that breaks the match); for ASCII symbols, this uses 8 bits.

In the absence of a match, the LZ77 token is (*0,0,sym*). Each LZ77 token uses $\log_2(|\text{dictionary}|) + \log_2(|\text{LAB}|) + 8$ bits; usually, |dictionary| ≫ |LAB|. In LZSS, the token has the format (*bit,code*),

with the structure of *code* depending on value *bit* as follows:

$$\begin{cases} bit = 0 & \Rightarrow \quad code = (sym), \\ bit = 1 & \Rightarrow \quad code = (pos,\ len). \end{cases} \quad (1)$$

In the absence of a match, LZSS produces (0,*sym*). The idea is that, when a match exists, there is no need to explicitly encode the next symbol. Besides this modification, Storer and Szymanski [16] also proposed keeping the LAB in a circular queue and the dictionary in a binary search tree, to optimize the search. LZSS is widely used in practice since it typically achieves higher compression ratios than LZ77. In LZSS, the token uses either 9 bits, when it has the form (0,*sym*), or $1 + \log_2(|\text{dictionary}|) + \log_2(|\text{LAB}|)$ bits, when it has the form (1,(*pos*,*len*)). The fundamental and most expensive component of these encoding algorithms is the search for the longest match between LAB prefixes and the dictionary.

Assuming that the decoder and encoder are initialized with equal dictionaries, the decoding of each LZ77 token (*pos*,*len*,*sym*) proceeds as follows: 1) *len* symbols are copied from the dictionary to the output, starting at position *pos* of the dictionary; 2) the symbol *sym* is appended to the output; 3) the string just produced at the output is slid into the dictionary. For LZSS decoding, we have: 1) if the bit field is 1, *len* symbols, starting at position *pos* of the dictionary, are copied to the output; otherwise *sym* is copied to the output; 2) the string just produced at the output is slid into the dictionary.

Both LZ77 and LZSS decoding are low complexity procedures, and thus decoding is much faster than encoding, because it involves no search. In this work, we address only the encoder side data structures and algorithms, with no effect in the decoder.

## 2.1 Suffix Arrays

A *suffix array* (SA) is the lexicographically sorted array of the suffixes of a string [7, 10]. For a string $D$ of length $m$ (with $m$ suffixes), the suffix array $P$ is the set of integers from 1 to $m$, sorted by the lexicographic order of the suffixes of $D$. For instance, if we consider dictionary $D = mississippi$ (with $m = 11$), its SA is $P = \{11, 8, 5, 2, 1, 10, 9, 7, 4, 6, 3\}$ and we get the suffixes shown in Fig. 1, along with the use of SA for LZ77/LZSS encoding:

- with $LAB = issia$, the LZ77 encoder outputs $(5, 4, a)$ or $(2, 4, a)$, depending on how we search $P$ and how we choose the match; for LZSS, we have $(1(5, 4))$ or $(1(2, 4))$ followed by $(0, (a))$;

- with $LAB = bsia$, the LZ77 tokens are $(0, 0, b)$ followed by $(7, 2, a)$ or $(4, 2, a)$; LZSS produces $(0, (b))$ followed by $(1(7, 2))$ or $(1(4, 2))$ and finally $(0, (a))$.

Each of the integers in $P$ is the suffix number corresponding to its position in $D$. Finding a sub-string of $D$ as in LZ77/LZSS, can be done by searching array $P$; for instance, the set of sub-strings of $D$ that start with symbol 's', can be found at indexes 7, 4, 6, and 3 of $D$, ranging from index 7 to 10 on $P$. There are several linear time algorithms for SA construction [8, 12, 18]; we have used the *suffix array induced sorting* (SA-IS) algorithm [12] .

## 3 Proposed Algorithm

We have adopted the following encoded file format. The header has 48 bits: the first 8 bits (*np*) represent the number of bits used to represent the *pos* field of the token; these are followed by another 8 bits ($nl$) with the number of bits used by the *len* field; the following 32 bits are the original file size. The header is followed by |LAB| ASCII symbols and the remainder of the file consists in a sequence of LZSS tokens. Our decoding algorithm does not need any special data structure and follows standard LZSS decoding, as described in Section 2.

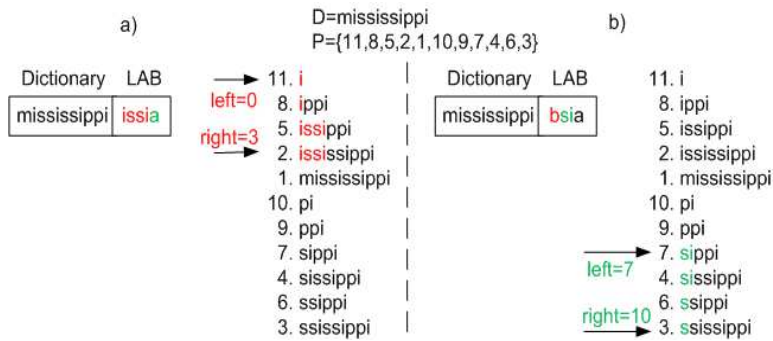

Figure 1: LZ77 and LZSS encoding with SA, with dictionary $D = mississippi$. In part a), with $LAB = issia$, we have four possible matches delimited by *left* and *right*. In part b) with $LAB = bsia$ there is no suffix that starts with 'b' (which is encoded as a single symbol), but after 'b' we find four suffixes whose first symbol is 's'; two of these suffixes start with 'si'.

The encoding algorithm uses two SA to represent the dictionary and an auxiliary array of 256 integers named LI (*LeftIndex*). This array holds, for each ASCII symbol, the first index of the suffix array where we can find the first suffix that starts with that symbol (the *left* index for each symbol, as shown in Fig. 1). The symbols such that are not the start of any suffix, the corresponding entry is marked with -1, meaning that we have an empty match for those symbols. Fig. 2 shows the LI for dictionary $D = mississippi$; for instance, the first suffix that starts with symbol 'i' is at index 0 in $P$, suffixes starting with 'p' are at index 5 of $P$. Using LI, we don't have to search for the *left* index for each sub-string in the LAB that we need to encode.

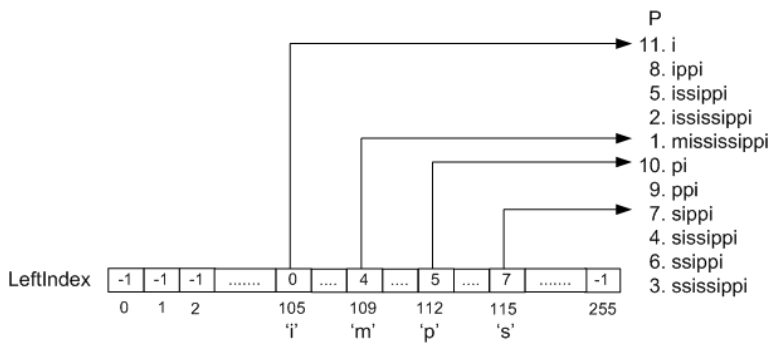

Figure 2: The LI (*LeftIndex*) auxiliary array: for each symbol that starts a suffix it holds the index of the SA $P$ in which that suffix starts. For the symbols that are not the start of any suffix, the corresponding entry is marked with -1, meaning that we have an empty match for sub-strings that start with that symbol.

As shown in Algorithm 1, the encoder starts by reading input symbols into the LAB. The first |LAB| symbols are written directly on the output file, because the dictionary is empty at that stage. We then slide the LAB into the dictionary and proceed by computing the SA for the dictionary until it is not full. When the dictionary is full, on every subsequent iteration, after each full LAB encoding, the corresponding SA and LI indexes are updated. This update,

**Algorithm 1** LZSS Encoding using Suffix Array
___
**Input:** $In$, input stream to encode;   $m$, length of dictionary;   $n$, length of LAB.
**Output:** $Out$, output stream with LZSS description of $In$.
 1: Write 48-bit header: $np$, $nl$ and $FileSize$ (as described above).
 2: Read the first look-ahead-buffer $LAB$, with |LAB| symbols, from $In$.
 3: Write LAB into $Out$.
 4: Initialize every position of LI to -1.
 5: Do coded $\leftarrow 0$.
 6: **while** $coded < FileSize$ **do**
 7:    Slide in LAB into dictionary $D$ and read next $LAB$.
 8:    **if** $coded < m$ **then**
 9:       Build SA, using SA-IS algorithm [12], for $D$ and name it $P$.        {/* Dictionary is filling. */}
10:    **else**
11:       **Update $P$ (as in algorithm UD and Fig. 3).**        {/* Runs after each LAB encoding. */}
12:    **end if**
13:    Scan $P$ and update LI (as described in Fig. 2).
14:    Do $i \leftarrow 0$.
15:    **while** $i < n$ **do**
16:       $left = LI[LAB[i]]$.                    {/* Loop to encode $n$ symbols in the $LAB$. */}
17:       **if** $(left == -1)$ **then**
18:          output $(0, LAB[i])$; $i \leftarrow i+1$; continue.   {/* Empty Match. No suffix starts with $LAB[i]$.*/}
19:       **end if**
20:       Find $right$, such that $D[P[right]] = LAB[i]$.        {/* Get *left* and *right* as in Fig. 1. */}
21:       From the set of suffixes between $P[left]$ and $P[right]$, choose the suffix at index $pos$, such that $left \leq pos \leq right$.        {/* Choose between "fast" and "best" compression. */}
22:       Do $len \leftarrow$ the match-length of sub-strings starting at $D[P[pos]]$ and $LAB[i]$.
23:       Output $(1(pos, len))$ into $Out$; $i \leftarrow i + len$.
24:    **end while**
25:    coded = coded + n.
26: **end while**
___

described in Algorithm 2 (named UD), runs when the dictionary is full, after each LAB encoding, performing the following actions: remove from $P$ the suffixes in the range $\{1, \ldots, |\text{LAB}|\}$ because they *slide out*; update in $P$ the suffixes in the range $\{|\text{LAB}| + 1, \ldots, |\text{dictionary}|\}$ to the range $\{1, \ldots, |\text{dictionary}| - |\text{LAB}|\}$, subtracting |LAB| to each suffix number; insert into $P$ the *slide in* suffixes in the range $\{|\text{dictionary}| - |\text{LAB}|, \ldots, |\text{dictionary}|\}$, after their proper sorting; this sorting is done by computing an SA for the LAB. To perform these actions on a single array is time consuming. To speed-up the update we use two SA of length |dictionary|, named $P_A$ and $P_B$, and a pointer $P$ (to $P_A$ or $P_B$). After each LAB encoding, we toggle pointer $P$ between $P_A$ and $P_B$, to avoid unnecessary removals, copies, and (slow) displacement of the elements of the working SA. After the update procedure, $P$ points to the new updated array. If the previous LAB encoding was done with $P_A$, the following will be carried out using $P_B$ and vice-versa. Fig. 3 illustrates this procedure with (|dictionary|,|LAB|)=(16,4) and the dictionary contents `this is the file`, with pointer $P$ set to $P_A$. We compute the SA for the LAB=`_the` and insert the new suffixes at indexes $\{2, 4, 8, 14\}$ of $P_B$; all other positions of $P_B$ are updated from $P_A$, subtracting |LAB| from $P_A$. After encoding LAB=`enco`, the update process is repeated using $P_A$ as destination.

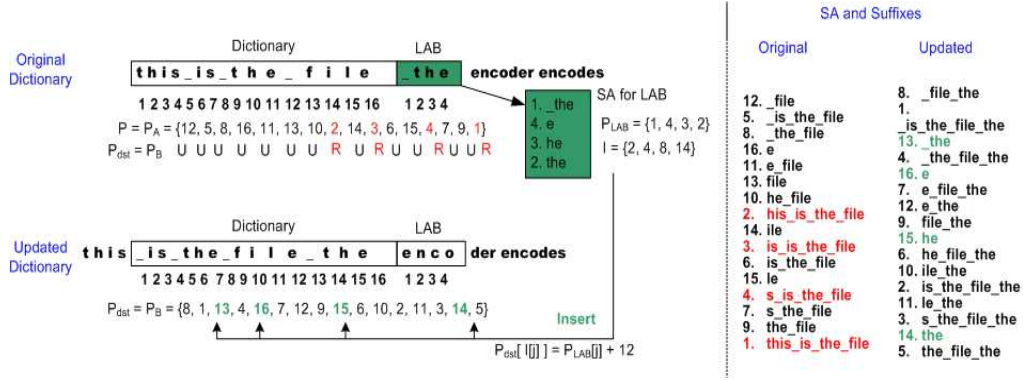

Figure 3: Update step with pointer $P$ set to $P_A$ initially; the first update is done using $P_B$ as destination. Array $I$ holds the indexes where to insert the new suffixes; 'U' and 'R' are the Update and Remove indexes, respectively. On the right hand side, we have the initial and updated dictionary contents.

---

**Algorithm 2** UD - Update Dictionary (line 11 of Algorithm 1)

---

**Input:** $P_A, P_B$, $m$-length SA;  $P$, pointer to PA or PB;  $P_{dst}$, pointer to PB or PA;
 $LAB$, look-ahead buffer;  LI, 256 position length LeftIndex array.
**Output:** $P_A$ or $P_B$ updated;  $P$ pointing to the recently updated SA.

1: **if** $P$ points to $P_A$ **then**
2:   Set $P_{dst}$ to $P_B$.
3: **else**
4:   Set $P_{dst}$ to $P_A$.
5: **end if**
6: Compute the SA $P_{LAB}$ for the encoded LAB.              {/* Sorts the suffixes in the LAB. */}
7: Using $LI$ and $P_{LAB}$, fill the $|LAB|$-length array $I$ with the insertion indexes (*slide in* suffixes).
8: **for** $j = 0$ to $|LAB| - 1$ **do**
9:   $P_{dst}[\ I[j]\ ] = P_{LAB}[j] + |dictionary| - |LAB|$.        {/* The $|LAB|$ Insert Suffixes. */}
10: **end for**
11: Do updateCounter $= |dictionary| - |LAB|$.
12: **for** $j = 0$ to $|dictionary| - 1$ **do**
13:   **if** $(P[j] - |LAB|) > 0$ **then**
14:     $P_{dst}[j] = P[j] - |LAB|$.              {/* The $|dictionary| - |LAB|$ Update Suffixes. */}
15:     updateCounter = updateCounter - 1;
16:     **if** (updateCounter==0) **then**
17:       break;        {/* Break immediately if $|dictionary| - |LAB|$ updates have been done. */}
18:     **end if**
19:   **end if**
20: **end for**
21: Set $P$ to $P_{dst}$.              {/* P points to recently updated SA. */}

---

In line 6 of Algorithm 2, we get the sorted suffixes corresponding to the recently encoded LAB. After line 7, array $I$ contains |LAB| integers with the indexes where these new suffixes are to be inserted (with lexicographical order) into $P_{dst}$; this search finds |LAB| positions, being quite fast because we use array LI to get the index of $P$, in which to start searching. The loop in lines 8 to 10 performs the sorted insertion at the corresponding indexes given by $I$ on the target SA pointed by $P_{dst}$. The loop in lines 12 to 20 updates the suffixes in the range $\{|\text{LAB}| + 1, \ldots, |\text{dictionary}|\}$ to the range $\{1, \ldots, |\text{dictionary}| - |\text{LAB}|\}$. With the use of two SA, we don't have to explicitly (slowly) remove the suffixes from the old SA.

We have also developed another version of Algorithm 1, which updates the SA at each and every token, thus being a *smooth sliding window suffix array*. The update procedure is divided into two situations, depending on the length of the token (1 or *len* symbols). For *len* symbols, we have the same procedure as described above, using *len* instead of |LAB|. When we have a match of a single symbol, we subtract one from each position of $P$, and remove the suffix corresponding to the single *slide out* symbol; finally, we insert the single suffix number |dictionary| corresponding to *symbol* at its corresponding position. This version turned out to be $2 \approx 3$ times slower than Algorithm 1, achieving about the same compression ratio.

## 4 Experimental Results

Our experimental tests were carried out on a laptop with a 2 GHz Intel Core2Duo T7300 CPU and 2 GB of RAM, using a single core. The code[1] was written in *C*, using Microsoft Visual Studio 2008. The linear time SA construction algorithm SA-IS [12] (available at http://yuta.256.googlepages.com/sais) was used. For comparison purposes, we also present the results of a BT-encoder [11], GZip[2], and the *LZ Markov chain algorithm* (LZMA[3]). The test files are from the standard corpora Calgary (18 files, 3 MB) and Silesia (12 files, 211 MB), available at http://www.data-compression.info. We use the "best" compression option (choice of the longest match, at line 21 of Algorithm 1).

### 4.1 Performance Indicators and Measures

In our tests, we used the Calgary and Silesia Corpus files, to assess the following measures: encoding time (in seconds, measured by the C function `clock`); compression ratio (in bits per byte, bpp); amount of memory for encoder data structures (in bytes). This amount for our encoder data structures is $M_{SA} = |\text{dictionary}| + |\text{LAB}| + 2|P| + |\text{LI}| + |\text{P\_LAB}|$. The integer array $LI$ has 256 positions, regardless of the length of the dictionary. $P_{LAB}$ is the SA for the LAB, with $|LAB|$ integers. The BT-encoder [11] uses 3 integers per tree node with $|\text{dictionary}| + 1$ nodes, occupying $M_{BT} = 13 \times |\text{dictionary}| + 12$ bytes, using 4-byte integers. A suffix tree algorithm[4] uses 3 integers and a symbol for each node, occupying 16 bytes, placed in a hash table [9], using the maximum amount of memory $M_{ST} = 25 \times |\text{dictionary}| + 4 \times \text{hashsz} + 16$ bytes, where *hashsz* is the hash table size. The GZip encoder occupies $M_{GZIP}=313408$ bytes, as measured with the 'C' `sizeof` operator. The LZMA encoder data structures occupy[5]

$$M_{LZMA} = 4194304 + \begin{cases} 9.5|\text{dictionary}|, & \text{if MF = BT2} \\ 11.5|\text{dictionary}|, & \text{if MF = BT3} \\ 11.5|\text{dictionary}|, & \text{if MF = BT4} \\ 7.5|\text{dictionary}|, & \text{if MF = HC4} \end{cases}, \quad (2)$$

---

[1]Available at http://www.deetc.isel.ipl.pt/sistemastele/docentes/AF/AF.htm
[2]http://www.gzip.org/
[3]http://www.7-zip.org
[4]http://www.larsson.dogma.net/research.html
[5]As reported in http://mancubus.net/svn/hosted/gzdoom/trunk/lzma/lzma.txt

Table 1: Amount of memory, total encoding time (in seconds), and average compression ratio (in bpb), for several lengths of (|dictionary|, |LAB|) on the Calgary Corpus, using "best" compression. GZip "fast" obtains Time=0.5 and bpb=3.20 while GZip "best" does Time=1.2 and bpb=2.79. The best encoding time is underlined.

| Calgary Corpus | | | SA "best" | | | BT "best" | | | LZMA "best" | | |
|---|---|---|---|---|---|---|---|---|---|---|---|
| # | |Dictionary| | |LAB| | Memory | Time | bpb | Memory | Time | bpb | Memory | Time | bpb |
| 1 | 2048 | 1024 | 24576 | <u>2.2</u> | 5.77 | 26636 | 3.92 | 5.65 | 4217856 | 4.7 | 2.99 |
| 2 | 4096 | 1024 | 43008 | <u>2.5</u> | 5.40 | 53260 | 4.3 | 4.98 | 4241408 | 4.8 | 2.82 |
| 3 | 4096 | 2048 | 48128 | <u>2.4</u> | 5.75 | 53260 | 11.1 | 5.48 | 4241408 | 4.8 | 2.82 |
| 4 | 8192 | 2048 | 84992 | <u>3.8</u> | 5.49 | 106508 | 11.7 | 4.88 | 4288512 | 5.1 | 2.69 |
| 5 | 16384 | 256 | 149760 | 9.1 | 4.36 | 213004 | <u>4.5</u> | 4.12 | 4382720 | 5.2 | 2.61 |
| 6 | 32768 | 256 | 297216 | 18.4 | 4.31 | 425996 | <u>5.5</u> | 4.08 | 4571136 | 4.9 | 2.54 |
| 7 | 32768 | 1024 | 301056 | 11.1 | 4.86 | 425996 | <u>7.5</u> | 4.40 | 4571136 | 4.9 | 2.54 |
| 8 | 32768 | 2048 | 306176 | <u>9.5</u> | 5.16 | 425996 | 15.8 | 4.57 | 4571136 | 4.9 | 2.54 |

bytes, depending on the *match finder* (MF) used as well as on |dictionary| with BT# denoting binary tree with # bytes hashing and HC4 denoting hash chain with 4 bytes hashing. For instance, with (|dictionary|, |LAB|) = (65536, 4096) we have $M_{SA}$ =611328, $M_{BT}$ =851980, $M_{ST}$ =1900560, and $M_{LZMA}$=4816896 bytes. If we consider an application in which we only have a low fixed amount of memory, such as the internal memory of an embedded device, it may not be possible to instantiate a tree or a hash table based encoder.

The GZip and LZMA[6] encoders perform entropy encoding of the dictionary tokens achieving better compression ratio than our LZSS encoding algorithms. Like GZip, LZMA is built upon the deflate algorithm, being the default compression method of 7z format in the 7-Zip program. These encoders are useful as a benchmark comparison, regarding encoding time and the amount of memory occupied. For both compression techniques, we have compiled their C/C++ sources using the same compiler settings, as for our encoders.

The compression ratio of our encoders as well as that of the BT-encoder can be easily improved by entropy-encoding the tokens, like in GZip and LZMA. Our purpose is to focus only on the construction of the dictionary and searching over it, using less memory than the conventional solutions with trees and hash tables.

## 4.2 Comparison with other encoders

We encode each file of the two corpora and compute the total encoding time as well as the average compression ratio, for different configurations of (|dictionary|, |LAB|), using "best" compression option. Table 1 shows the results of these tests on the Calgary Corpus. Our SA-encoder is faster than BT, except on tests 5 to 7; on test 6 (the GZip-like scenario), BT-encoder is about 3.5 times faster than SA. Table 2 shows the results for the Silesia Corpus. In these tests, the SA-encoder is the fastest except on tests 5 and 6. On test 3, the SA-encoder is about 5 times faster than the BT-encoder, achieving about the same compression ratio. We see that when |LAB| is not too small (as compared to the dictionary), the SA-encoder is faster than the BT-encoder. Fig. 4 shows the trade-off between time and memory on the encoding of the Calgary and Silesia corpora, on the tests shown on Tables 1 and 2, for SA and BT-encoders, including GZip test results for comparison; the SA-encoder offers a good trade-off, especially on tests 1 to 5, using (much) less memory than GZip.

---

[6]LZMA SDK, version 4.65, released 3 February 2009, available at http://www.7-zip.org/sdk.html

Table 2: Amount of memory, total encoding time (in seconds) and average compression ratio (in bpb), for several lengths of (|dictionary|, |LAB|) on the Silesia Corpus, using "best" compression. GZip "fast" obtains Time=19.5 and bpb=3.32 while GZip "best" does Time=74.4 and bpb=2.98. The best encoding time is underlined.

| | Silesia Corpus | | SA "best" | | | BT "best" | | | LZMA "best" | | |
|---|---|---|---|---|---|---|---|---|---|---|---|
| # | \|Dictionary\| | \|LAB\| | Memory | Time | bpb | Memory | Time | bpb | Memory | Time | bpb |
| 1 | 2048 | 1024 | 24576 | 118.7 | 5.66 | 26636 | 249.5 | 5.65 | 4217856 | 333.53 | 3.05 |
| 2 | 4096 | 1024 | 43008 | 116.9 | 5.41 | 53260 | 303.4 | 5.25 | 4241408 | 349.05 | 2.90 |
| 3 | 4096 | 2048 | 48128 | 112.9 | 5.68 | 53260 | 694.9 | 5.63 | 4241408 | 349.05 | 2.90 |
| 4 | 8192 | 2048 | 84992 | 143.4 | 5.44 | 106508 | 668.9 | 5.27 | 4288512 | 356.77 | 2.76 |
| 5 | 16384 | 256 | 149760 | 319.1 | 4.55 | 213004 | 254.6 | 4.44 | 4382720 | 366.47 | 2.62 |
| 6 | 32768 | 256 | 297216 | 542.7 | 4.41 | 425996 | 318.1 | 4.31 | 4571136 | 356.34 | 2.52 |
| 7 | 32768 | 1024 | 301056 | 322.2 | 4.80 | 425996 | 382.6 | 4.64 | 4571136 | 356.34 | 2.52 |
| 8 | 32768 | 2048 | 306176 | 302.3 | 5.02 | 425996 | 979.8 | 4.81 | 4571136 | 356.34 | 2.52 |

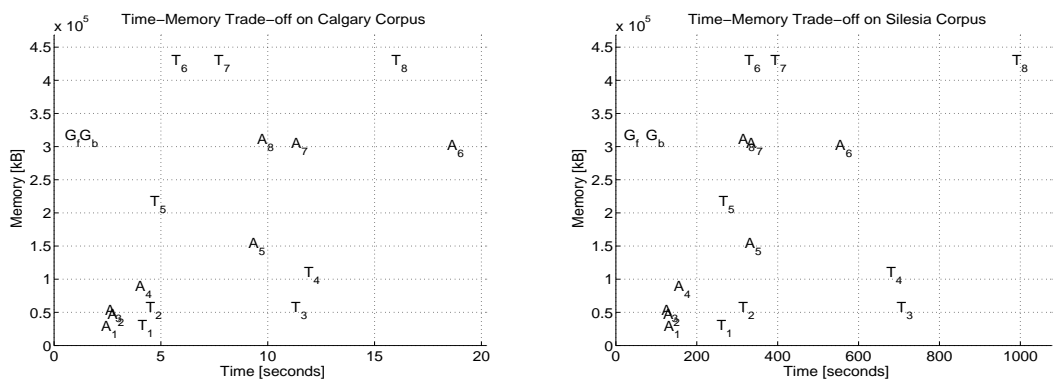

Figure 4: Time-memory trade-off between SA (A#) and BT (T#) for the Calgary and Silesia Corpus, on the 8 encoding tests of Tables 1 and 2. We include GZip for comparison: $G_f$ is GZip "fast", $G_b$ is GZip "best".

For all these encoders searching and updating the dictionary are the most time-consuming tasks. A high compression ratio like those of LZMA and GZip can be attained only when we use entropy encoding with appropriate models for the tokens. The SA encoder is faster than the BT encoder, when the LAB is not too small. Our algorithms (without entropy encoding) are thus positioned in a trade-off between time and memory, that can make them suitable to replace binary trees on LZMA or in sub-string search.

## 5 Conclusions

In this paper, we have proposed a new Lempel-Ziv encoding algorithm based on suffix arrays, improving on earlier work in terms of encoding time, being faster than previous approaches, with similar low memory requirements. The proposed algorithm uses an auxiliary array as an accelerator to the encoding procedure, as well as a fast update of the dictionary based on two suffix arrays. This algorithm has considerably lower memory requirements than binary/suffix trees and hash tables.

The proposed algorithm allows *a priori* computing the exact amount of memory necessary for the encoder data structures; usually this may not be the case when using (binary/suffix) trees, because the number of nodes and branches to allocate depends on the contents of the text, or when we allocate a memory block that is larger than needed as it happens with hash tables.

We have compared our algorithm (on benchmark files from standard corpora) against tree-based encoders, including GZip and LZMA. The experimental tests showed that in some (typical) compression settings, our encoders occupy less memory and are faster than tree-based encoders, thus being time *and* memory efficient LZ77 and LZSS encoders, based on suffix arrays. The tree-based encoders can only be faster at the expense of memory usage. Our algorithm is positioned in a trade-off between time and memory, that can make it suitable to replace the use of trees, like in LZMA (7-Zip), reducing the amount of memory and encoding time, keeping the same compression ratio, which is better than that of GZip.

These encoders also provide a more compact way to represent the dictionary which is suited for text categorization, based on bag-of-words representations. Using a single suffix array and the 256-position auxiliary array, we have a fast indexing mechanism to quickly find all the sets of words that start with a given symbol, on a static dictionary. This will be topic of future research.